\documentstyle[aps,prb,preprint]{revtex}

\begin{document}

\draft

\title{Step edge barriers on GaAs(001)}
\author{Pavel \v{S}milauer\cite{byline1,byline2}}
\address{Interdisciplinary Research Centre for Semiconductor Materials,
Imperial College, London SW7~2BZ, United Kingdom}
\author{Dimitri D. Vvedensky}
\address{Department of Physics, The Blackett Laboratory, Imperial College,
London SW7~2BZ, United Kingdom}

\date{\today}

\maketitle

\begin{abstract}

We investigate the growth kinetics on vicinal GaAs(001) surfaces by making
detailed comparisons between reflection high--energy electron--diffraction
specular intensity measured near in--phase diffraction conditions and
the surface step density obtained from simulations of a solid--on--solid
model. Only by including a barrier to interlayer transport and a
short--range incorporation process of freshly--deposited atoms can the
simulations be brought into agreement with the measurements both during
growth and during post--growth equilibration of the surface.

\end{abstract}


The roughness of growing surfaces has become one of the main topics in the
study of surface processes. One reason for this is the evident importance
of
minimizing surface roughness for applications. Aside from its relevance for
the fabrication of narrow layered structures (quantum wells, lateral
superlattices, magnetic multilayers, etc.), surface roughness has recently
been suggested as triggering the transition from epitaxial to amorphous
growth during Si, Ge, and GaAs homoepitaxy.\cite{egc} Provided the
roughness can be controlled, this opens up the exciting prospect of
low--temperature epitaxy and doping. Another reason for the intense
interest in surface roughness is purely theoretical. The roughness of
growing surfaces has been observed to exhibit asymptotic dynamical scaling
behavior\cite{fam} which has led to the classification of growth models
into various universality classes.

In a majority of the theoretical studies, the focus has been on the
evolution of the surface morphology due to fluctuations in the incoming
flux
of particles and surface diffusion. Studies of metal epitaxy have
revealed the importance of two processes which directly influence the
evolution of surface roughness. The first is the way  freshly--deposited
atoms are incorporated into the growing material  (``transient
mobility'',\cite{ej} ``downward funneling'',\cite{ev} etc.). Such
incorporation mechanisms smoothen the growth front and lead to quasi
layer--by--layer growth at temperatures low enough that thermal mobility is
negligible.\cite{ej,ev,klb,sn,f,swv} The second process is the interlayer
transport of material and, in particular, the effect of activation barriers
to adatom hopping between layers. These barriers lead to the opposite trend
of the incorporation just described, namely, a rapid roughening of the
growing {\it singular\/} surface.\cite{f,swv,kpv,egj,vill}

While the effects of an incorporation process and hopping barriers are
clearly manifested on metal surfaces, the question arises as to whether
these processes might also play an observable role in the growth of
semiconductor surfaces.  In this paper we address this question by
investigating the growth kinetics on vicinal GaAs(001) surfaces. We make
direct comparisons between the evolution of the measured reflection
high--energy electron--diffraction (RHEED) specular--beam intensity and the
step density of simulated surfaces. RHEED is an {\it in situ\/} real--time
probe of surface morphology and, as numerous studies have
shown,\cite{rheed} can be a sensitive measure of the presence of various
surface processes.  We find that {\it quantitative\/} agreement between the
RHEED specular intensity and the step density can be achieved {\it both\/}
during growth and post--growth equilibration of the surface if our model
includes barriers to interlayer transport and an incorporation process for
arriving atoms.

We first describe the basic simulation model\cite{cv} before including the
refinements just described. The growing crystal is assumed to have a simple
cubic structure with neither vacancies nor overhangs (the solid--on--solid
model\cite{wg}). Growth is initiated by the random deposition of atoms
onto the substrate at a rate determined by the flux. The subsequent
migration of surface adatoms is taken as a nearest--neighbor hopping
process with the rate $k(T)\!=\!k_0\exp(-E_D/k_BT)$, where $T$ is the
substrate temperature, $k_0\!=\!2k_BT/h$, $k_B$ is Boltzmann's constant,
$h$ is Planck's constant and $E_D$ is the hopping barrier. The latter is
comprised of a term, $E_S$, from the substrate, and a contribution, $E_N$,
from each lateral nearest neighbor. Thus, the hopping barrier of an atom
with $n$ lateral nearest neighbors is $E_D\!=\!E_S\!+\!nE_N$, where
$n\!=\!0,\ldots,4$. Thermal desorption is neglected.

In attempting to develop a model with as few free parameters as possible
while retaining the essential features of the growth kinetics, some
simplifying assumptions have been introduced. Those that require the
closest examination are as follows. (i) The mobility and nearest--neighbor
bonding of adatoms are isotropic. (ii) The group V kinetics are not
included explicitly in the model, since it is assumed that under normal
growth conditions the group V species is present in sufficient quantities
to insure microscopic stoichiometry. (iii) The effects of the surface
reconstruction on mobility can be incorporated as part of the effective
migration parameters ($E_S$ and $E_N$).

In the experiments reported by Shitara {\it et al.\/},\cite{s1} growth and
diffraction conditions were chosen to conform as closely as possible to
these assumptions. In particular, to satisfy (i) the surfaces were
misoriented toward the [010] direction to reduce the effect of the
anisotropy, to satisfy (ii) the As/Ga (atomic) ratio was held at
approximately 2.5 to maintain the 2$\times$4 reconstruction in a fairly
wide temperature range near the temperature at which growth becomes
dominated by step flow. This also addresses assumption (iii).  For the
reasons discussed in Ref.~\onlinecite{s1}, the diffraction conditions were
chosen as ``in--phase'', or ``Bragg'' conditions for which the maxima in
the
specular--beam intensities on the corresponding singular surface occur at
monolayer (ML) increments of material deposited.  Since for in--phase
diffraction conditions the kinematic theory yields a constant intensity,
regardless of the surface configuration, the density of surface steps was
used to model the {\it variations\/} of the RHEED specular--beam intensity
during growth.  The most striking result to emerge from this study is the
extent of {\it quantitative\/} agreement between the RHEED specular--beam
intensity and the step density evolutions during growth to the extent that
the relative changes of the magnitudes of the two quantities with
temperature are the same.

On the other hand, there was also clear disagreement during post--growth
recovery (smoothing of the surface), which was most evident at lower
substrate temperatures. The simulations generally showed a much more  rapid
recovery than the measured RHEED profiles.\cite{s1} More important, these
simulations were unable to reproduce the systematic dependencies of the
time constant $\tau_1$, Ref.~\onlinecite{recovery}, for the initial stage
of
the recovery (cf.~Refs.~\onlinecite{vc} and \onlinecite{s3}). These are
serious discrepancies because the equilibration of a surface after a period
of growth is a more discriminating test of kinetic models than growth
alone.
During growth under typical conditions, the maximum time scale is set by
the deposition flux and processes occurring over longer time scales are
``frozen out''. However, during recovery, these processes can come into
play even though their effect during growth can be safely omitted.\cite{vc}

To more accurately model the equilibration of the surface, we introduce an
activation barrier to hopping at step edges, $E_B$, of the same form as
that used in our simulations of the reentrant oscillations seen in
thermal--energy atom scattering measurements on Pt(111).\cite{swv} This
barrier is expected to provide a strong signature in the recovery profile
by
inhibiting the interlayer transport of adatoms and thus slowing down the
recovery.\cite{pw}  However, even a quite small barrier to interlayer
hopping has important consequences during growth as well. The surface
quickly roughens and complete disagreement between simulations and
experiment during growth {\it and\/} recovery is observed. We are thus led
to the conclusion that if step--edge barriers are present then some
compensating smoothing mechanism must also be present. We have therefore
included an additional process that an arriving atom undergoes before being
incorporated on the surface. A search is carried out within a square of a
fixed linear size $L$, centered upon the original site, for the site that
maximizes the number of nearest neighbors. The effect of this process is a
smoothing of edges of pre--existing steps and islands created on the
terraces.  A similar mechanism has been invoked to explain the monolayer to
bilayer transition in the growth of group IV materials,\cite{cywv} though
we can only speculate about its possible microscopic origins for the case
of
GaAs(001).  The simulations reported below were carried out on
232$\times$232 lattices with steps running diagonally across the lattice (a
miscut toward [010] direction)\cite{s1} with the parameters
$E_S\!=\!1.54$eV, $E_N\!=\!0.23$eV, $E_B\!=\!0.175$eV and $L\!=\!7$.

A comparison of our modified model with the data of Ref.~\onlinecite{s1}
for two different Ga fluxes is shown in Figs.~\ref{rec47} and \ref{rec20}.
The step densities show a much better agreement with the RHEED curves over
the entire growth and recovery periods than those produced with the
original model. The incorporation process brings the simulations and the
measurements into closer agreement at the onset of growth and during
growth, where the improvement is manifested particularly in a much better
reproduction of the first maximum delay phenomenon.\cite{frst} This is due
to an increase in the number of atoms incorporated into pre-existing steps
immediately after they are deposited. In Fig.~\ref{tau} is an Arrhenius
plot of the temperature dependence of $\tau_1$, which now {\it does\/}
exhibit an Arrhenius dependence, in agreement with
measurements.\cite{recovery}

The comparisons in Figs.~\ref{rec47} and \ref{rec20} are compelling not
least because the effects of the incorporation process and step--edge
barrier act in {\it opposition\/} during growth, as discussed above, but in
{\it concert\/} during recovery.  The step--edge barrier slows down the
equilibration process by inhibiting interlayer transport while the
incorporation process produces a smoothing of step edges and a decrease in
the number of free adatoms which leads to a high coordination of most of
the adatoms forming clusters on the uppermost layer. This reduces the
influence of fast processes such as free adatoms migrating to coordinated
sites and the elimination of sites with low coordination, both of which
were
over--emphasized in the original model. Therefore, according to our
simulations, both step-edge barriers and the incorporation process are
important factors for correctly reproducing the recovery curves.

A question naturally arises:~are there any other experimental hints of
there
being barriers at step edges or an incorporation process on semiconductor
surfaces? The role of step--edge barriers has been examined in the
regularization of terrace width distributions on vicinal
surfaces.\cite{ctg} The existence of step--edge barriers has also been
suggested recently as being responsible for the unexpectedly rapid increase
in the surface roughness in low--temperature growth of Si
(Ref.~\onlinecite{egj}). The ``forbidden temperature window'' in the growth
of AlAs(001)\cite{dcc} is very suggestive of reentrant layer--by--layer
growth behavior caused by the step--edge barriers. Alternatively, the
observations of epitaxial growth at very low temperatures\cite{egc,m} could
be explained as the effect of an incorporation process.

A final comment is in order concerning the results presented here with
regard to the presence of As. Shitara {\it et al.\/}\cite{s1} observed that
the areas of disagreement between the RHEED measurements and the original
model are in transient regimes of growth before the As/Ga ratio has
attained a steady state value.\cite{s3} The explanation of the recovery
effect put forth in this paper relies on the value of the step--edge
barrier (and also other model parameters) being the same during growth and
recovery. In particular, if the step edges are the preferred sites for
adsorbed As, then maintaining a steady state As/Ga ratio over the entire
surface is not required, though the absence of As at these sites would be
expected to have an observable effect. In fact, at the highest
temperatures, where As desorption becomes most appreciable, the step
density recovers more slowly than the RHEED specular--beam intensity
(Figs.~\ref{rec47} and \ref{rec20}), suggesting that a step edge without As
has a lower barrier to interlayer hopping than the same edge with an
adsorbed As.

In conclusion, we have studied growth and post--growth recovery on
GaAs(001)
surfaces using computer simulations of a solid--on--solid model. To the
best
of our knowledge, we have achieved for the first time {\it quantitative\/}
agreement between the evolution of the RHEED specular--beam intensity and
its theoretical counterpart, in our case the surface step density. Our
results support strongly the idea of there being step--edge barriers to
interlayer transport as well as an incorporation process for arriving atoms
on GaAs(001) surfaces.

We would like to thank Dr.~T.~Shitara for providing us with his
experimental and simulation results and for stimulating discussions and
suggestions. P.\v{S}.~would like to thank Dr.~M.~Wilby for useful
discussions on the simulation model. The support of Imperial College and
the Research Development Corporation of Japan under the auspices of the
``Atomic Arrangement:~Design and Control for New Materials'' Joint Research
Program is gratefully acknowledged.

\begin{figure}
\caption{Direct comparison between measured RHEED intensity and the
simulated surface step density on a GaAs(001) vicinal surface
misoriented by $2^\circ$ toward [010] direction at the growth rate of
0.47~ML/s.}
\label{rec47}
\end{figure}

\begin{figure}
\caption{Direct comparison between measured RHEED intensity and the
simulated surface step density on a GaAs(001) vicinal surface misoriented
by $2^\circ$ toward [010] direction at the growth rate of 0.20~ML/s.}
\label{rec20}
\end{figure}

\begin{figure}
\caption{The Arrhenius temperature dependence of the time constant $\tau_1$
of the fast stage of post--growth recovery.}
\label{tau}
\end{figure}

\end{document}